\documentclass[a4paper,10pt]{article}
\usepackage{latexsym}
\usepackage{amsmath}
\usepackage{amssymb}
\usepackage{amscd}

\newlength{\minitwocolumn}
\makeatletter
\@addtoreset{equation}{section}
\makeatother

\newtheorem{thm}{Theorem}[section]
\newtheorem{prop}[thm]{Proposition}
\newtheorem{lem}[thm]{Lemma}

\title{\bf Type II vertex operators 
for the $A_{n-1}^{(1)}$ face model}

\author{H. Furutsu$^*$, T. Kojima$^*$ and 
Y.-H. Quano$^\dagger$}

\date{\it ${}^*$Department of Mathematices,
Colledge of Science and Technology, 
Nihon University, Chiyoda-ku, Tokyo 101-0062, Japan \\
\it ${}^\dagger$Department of Medical Electronics, 
Suzuka University of Medical Science \\
      \it Kishioka-cho, Suzuka 510-0293, Japan \\
\today}
\begin{document}

\maketitle
\begin{abstract}
Presented is a free boson representation of 
the type II vertex operators 
for the
$A_{n-1}^{(1)}$ face model. 
Using the bosonization, we derive some properties
of the type II vertex operators, such as 
commutation, inversion and duality relations. 
\end{abstract}
\section{Introduction}

Recent development on integrable lattice models 
is mainly based on the representation theory of the 
quantum affine groups \cite{XXZ,JM}. 
For example, the XXZ spin chain model has 
the $U_q (\widehat{\mathfrak s\mathfrak l}_2)$ 
symmetry, which 
is large enough to diagonalize the XXZ Hamiltonian. 
Correlation functions and form factors are also 
given as the traces of products of vertex operators. 
The integral formulae of the correlation 
functions for the XXZ model were presented by using 
the free boson realization of the vertex \cite{JM,CORR}. 

In \cite{XXZ} vertex operators 
were introduced as the intertwiners among the 
highest weight modules and the finite dimentional 
representation of quantum affine groups. 
There exist two kinds of vertex operators, 
the type I and the type II vertex operaors in the 
terminology of  \cite{XXZ}, 
because of the left-right assymmetry of the 
coproduct of the quantum affine groups. 
The XXZ Hamiltonian is 
expressed in terms of the type I vertex operators, 
whereas the creation and annihilation operators are 
constructed in terms of the type II 
vertex operators \cite{XXZ}. 

The vertex operator method can be also 
formulated for face models \cite{JMOh,FJMMN}. 
Bosonization of the type I vertex operators 
for the RSOS model \cite{ABF} were given by 
Lukyanov and Pugai \cite{LP}, in which they 
derived an integral representation 
for multi-point local state probabilities. 
In \cite{AJMP} a bosonization of the type I 
vertex operators was realized 
for the $A^{(1)}_{n-1}$ face 
model \cite{JMO}, 
that reduces to the RSOS model \cite{ABF} 
when $n=2$. In \cite{MW} a bosonization 
of the type II vertex operators was realized 
for the RSOS model. The 
$A^{(1)}_{n-1}$ face model has the defromed 
Virasoro algebra symmetry \cite{SKAO}. Since 
the Virasoro algebra has no non-trivial defromation 
of the coproduct, those vertex operators cannot 
be interpreted 
as intertwiners of the defromed Virasoro algebra. 
However, the type I vertex operators 
allow a graphical interpretation which is due to 
the identification between 
the bosonized vertex operators and the 
half transfer matrix for the $A^{(1)}_{n-1}$ face 
model \cite{LP,AJMP}. 

In this paper we present 
a bosonization of the type II vertex operators 
for the $A^{(1)}_{n-1}$ face model. Unfortunately, 
the type II vertex operators do not allow 
interpretation  
as intertwiners of the defromed Virasoro algebra 
nor graphical interpretation. Our construction 
is thus based on the commutation relations 
among the type I and II vertex operators. 

The rest of the paper is organized as follows. 
In section 2 we formulate the problem and introduce 
the type II vertex operators for the $A^{(1)}_{n-1}$ 
face model. In section 3 we realize the type II 
vertex operators for the $A^{(1)}_{n-1}$ face model
in terms of free bosons. We prove the commutation 
relations of the those bsonaized operators in order 
to show that they are bona fide vertex operators. 
In section 4 we prove various properties of the 
vertex operatos such as the inversion and the duality 
relations. 

\section{Type II vertex operators for 
the $A^{(1)}_{n-1}$ face model}

The present section aims to formulate 
the problem, thereby fixing the notation. 

\subsection{Theta functions} 

Throughout this paper we fix the integers $n$ and 
$r$ such that $r\geqq n+2$, and also fix 
the parameter $x$ such that $0<x<1$. 
We will use the abbreviations, 
\begin{equation}
[v]=x^{\frac{v^2}{r}-v}\Theta_{x^{2r}}(x^{2v}), 
~~~~
[v]'=x^{\frac{v^2}{r-1}-v}\Theta_{x^{2r-2}}(x^{2v}), 
\end{equation}
where the Jacobi theta function is given by 
\begin{eqnarray}
\Theta_{q}(z)&=&(z; q)_\infty 
(qz^{-1}; q)_\infty (q, q)_\infty , \\
(z; q_1 , \cdots , q_m )&=& 
\prod_{i_1 , \cdots , i_m \geqq 0} 
(1-zq_1^{i_1} \cdots q_m^{i_m}). 
\end{eqnarray}

For later conveniences we also 
introduce the following symbols
\begin{eqnarray}
r_{l}(v)&=&z^{\frac{r-1}{r}\frac{n-l}{n}}
\frac{g_{l}(z^{-1})}{g_{l}(z)}, ~~~~
g_{l}(z)=
\frac{\{x^{2n+2r-l-1}z\}
\{x^{l+1}z\}}
{\{x^{2n-l+1}z\}\{x^{2r+l-1}z\}}, \\
r^*_{l}(v)&=&z^{\frac{r}{r-1}\frac{n-l}{n}}
\frac{g^*_{l}(z^{-1})}{g^*_{l}(z)}, ~~~~
g^*_{l}(z)=
\frac{\{x^{2n+2r-l-1}z\}'
\{x^{l-1}z\}'}
{\{x^{2n-l-1}z\}'\{x^{2r+l-1}z\}'}, \\
\chi_l (v)&=&z^{\frac{l(l-n)}{n}}
\frac{\rho_l (z^{-1})}{\rho_l (z)}, ~~~~
\rho_l (z)=
\frac{(-x^{2l+1}z;x^{2n},x^2)_\infty
(-x^{2n-2l+1}z;x^{2n},x^2)_\infty
}{(-xz;x^{2n},x^2)_\infty
(-x^{2n+1}z;x^{2n},x^2)_\infty
}
\end{eqnarray}
where $z=x^{2v}$, $1\leqq l\leqq n$ and 
\begin{equation}
\{z\}=(z;x^{2r},x^{2n})_\infty , 
~~~~
\{z\}'=(z;x^{2r-2},x^{2n})_\infty . 
\label{eq:{z}}
\end{equation}
In particular we define $R(v), 
S(v)$ and $\chi (v)$ by 
\begin{eqnarray}
R(v)=r_1 (v)=
z^{\frac{r-1}{r}\frac{n-1}{n}}
\frac{\rho(z^{-1})}{\rho(z)}, && 
\rho(z)=g_1 (z)
=\frac{\{x^2z\} \{x^{2n+2r-2}z\}}
{\{x^{2r}z\}\{x^{2n}z\}},
\label{def:R} \\
S(v)=s_1 (v)=
z^{-\frac{r}{r-1}\frac{n-1}{n}}
\frac{\rho^{*}(z^{-1})}
{\rho^{*}(z)}, && 
\rho^{*}(z)=g^*_1 (z)=
\frac{\{x^{2n-2}z\}'\{x^{2r}z\}'}
{\{x^{2n+2r-2}z\}'\{z\}'}, 
\label{def:S}
\end{eqnarray}
and
\begin{equation}
\chi(v)=\chi_1 (v)=
z^{\frac{1-n}{n}}
\frac{(-x^{2n-1}z^{-1};x^{2n})_\infty 
(-xz ;x^{2n})_\infty}
{(-xz^{-1};x^{2n})_\infty (-x^{2n-1}z; x^{2n})_\infty}. 
\label{def:chi}
\end{equation}
These facotrs will appear in the commutation relations 
among the type I and the type II 
vertex operators. 

The integral kernel for the type I and the type II 
vertex operators will be given as the products of 
the following elliptic functions 
\begin{eqnarray}
f(v,w)=\frac{[v+\frac{1}{2}-w]}{[v-\frac{1}{2}]},~~~~
h(v)=\frac{[v-1]}{[v+1]},\\
f^*(v,w)=\frac{[v-\frac{1}{2}+w]'}{[v+\frac{1}{2}]'},~~~
h^*(v)=\frac{[v+1]'}{[v-1]'}. 
\end{eqnarray}

\subsection{The weight lattice of $A^{(1)}_{n-1}$}

Let $V=\mathbb{C}^n$ and 
$\{ \varepsilon _\mu \}_{1 \leqq j \leqq n}$ be 
the standard orthogonal basis with the inner 
product $\langle \varepsilon _\mu , 
\varepsilon _\nu \rangle =\delta_{\mu \nu}$. 
The weight lattice of $A^{(1)}_{n-1}$ 
is defined as follows: 
\begin{equation}
P=\bigoplus_{\mu =1}^n 
\mathbb{Z} \bar{\varepsilon}_\mu , 
\label{eq:wt-lattice}
\end{equation}
where 
$$
\bar{\varepsilon}_\mu =
\varepsilon _\mu -\varepsilon , 
~~~~\varepsilon =\frac{1}{n}\sum_{\mu =1}^{n} 
\varepsilon _\mu . 
$$
We denote the fundamental weights 
by $\omega_\mu\,(1\leqq \mu \leqq n-1)$ 
$$
\omega_\mu =\sum_{\nu =1}^\mu 
\bar{\varepsilon }_\nu , 
$$
and also denote the simple roots by 
$\alpha_\mu \,(1\leqq \mu \leqq n-1)$ 
$$
\alpha_\mu =\varepsilon _\mu -
\varepsilon _{\mu +1}=\bar{\varepsilon}_\mu 
-\bar{\varepsilon}_{\mu +1}. 
$$
For $a\in P$ we set 
\begin{equation}
a_{\mu\nu}=\langle a+\rho , 
\varepsilon _\mu -\varepsilon _\nu \rangle , ~~~~ 
\rho =\sum_{\mu =1}^{n-1} \omega_\mu . 
\end{equation}

\subsection{The $A^{(1)}_{n-1}$ face model}

An ordered pair $(a,b) \in P^2$ 
is called {\it admissible} if $b=a+\bar{\varepsilon}_\mu$, 
for a certain $\mu\,(1\leqq \mu \leqq n)$. 
For $(a, b, c, d)\in P^4$ let 
$\displaystyle W_I 
\left( \left. \begin{array}{cc} 
c & d \\ b & a \end{array} 
\right| v \right) $ 
be the Boltzmann weight of the 
$A^{(1)}_{n-1}$ model for the state configuration 
$\displaystyle 
\left( \begin{array}{cc} 
c & d \\ b & a \end{array} \right) $ 
round a face. 
Here the four states $a, b, c$ and $d$ are 
ordered clockwise from the SE corner. 
In this model $W_I 
\left( \left. \begin{array}{cc} 
c & d \\ b & a \end{array} \right| 
v \right) =0~~$ 
unless the four pairs $(a,b), (a,d), (b,c)$ 
and $(d,c)$ are admissible. 
Non-zero Boltzmann weights are parametrized in terms of 
the elliptic theta function of the spectral parameter $v$ 
as follows: 
\begin{equation}
\begin{array}{rcl}
W_{I}
\left( \left. \begin{array}{cc} 
a + 2 \bar{\varepsilon }_\mu & a+\bar{\varepsilon }_\mu \\ 
a+\bar{\varepsilon }_\mu & a \end{array} \right| 
v  \right) 
& = & R(v), \\
~ & ~ & ~ \\
W_{I}
\left( \left. \begin{array}{cc} 
a+\bar{\varepsilon }_\mu +\bar{\varepsilon }_\nu 
& a+\bar{\varepsilon }_\mu \\ 
a+\bar{\varepsilon }_\nu & a 
\end{array} \right| v \right) & = & R(v)
\dfrac{[v][a_{\mu\nu}-1]}{[v-1][a_{\mu\nu}]} 
~~~~(\mu \neq \nu ), \label{BW} \\
~ & ~ & ~ \\
W_{I}
\left( \left. \begin{array}{cc} 
a+\bar{\varepsilon }_\mu +\bar{\varepsilon }_\nu 
& a+\bar{\varepsilon }_\nu \\ 
a+\bar{\varepsilon }_\nu & a 
\end{array} \right| v \right) & = & R(v)
\dfrac{[1][v-a_{\mu\nu}]}{[v-1][a_{\mu\nu}]} 
~~~~(\mu\neq \nu). 
\end{array}
\end{equation}

The Boltzmann weights (\ref{BW}) 
solve the Ynag-Baxter equation 
for the face model \cite{JMO}: 
\begin{equation}
\begin{array}{cc}
~ & 
\displaystyle \sum_{g} 
W_{I}\left( \left. 
\begin{array}{cc} d & e \\ c & g \end{array} \right| 
v_1 \right)
W_{I}\left( \left. 
\begin{array}{cc} c & g \\ b & a \end{array} \right| 
v_2 \right)
W_{I}\left( \left. 
\begin{array}{cc} e & f \\ g & a \end{array} \right| 
v_1 -v_2 \right) \\
~ & ~ \\
= & \displaystyle \sum_{g} 
W_{I}\left( \left. 
\begin{array}{cc} g & f \\ b & a \end{array} \right| 
v_1 \right)
W_{I}\left( \left. 
\begin{array}{cc} d & e \\ g & f \end{array} \right| 
v_2 \right)
W_{I}\left( \left. 
\begin{array}{cc} d & g \\ c & b \end{array} \right| 
v_1 -v_2 \right)
\end{array} \label{YBE}
\end{equation}

\subsection{Commutation relations}

Consider the type I vertex operators satisfying 
the following commutation relations 
\begin{eqnarray}
{\Phi}_{\mu_1}(v_1)\Phi_{\mu_2}(v_2)
=\sum_{\varepsilon_{\mu_1}+\varepsilon_{\mu_2}
=\varepsilon_{\mu_1'}+\varepsilon_{\mu_2'} }W_{I}\left(\left.
\begin{array}{cc}
k+\bar{\varepsilon}_{\mu_1}+\bar{\varepsilon}_{\mu_2}&
k+\bar{\varepsilon}_{\mu_2'}\\
k+\bar{\varepsilon}_{\mu_1}&k
\end{array}\right|v_1-v_2 \right)
{\Phi}_{\mu_2'}(v_2)
{\Phi}_{\mu_1'}(v_1). 
\label{eq:CR-I}
\end{eqnarray}
The bosonization of ${\Phi}_{\mu}(v)$ is 
realized in \cite{AJMP}. See also section 3. 
The dual type II vertex operators 
should obey the following commutation relations 
\begin{eqnarray}
{\Phi}_{\mu_1}(v_1)\Psi_{\mu_2}^*(v_2)
=\chi(v_1-v_2)\Psi_{\mu_2}^*(v_2)
{\Phi}_{\mu_1}(v_1), 
\label{bp2}
\end{eqnarray}
\begin{eqnarray}
\Psi_{\mu_1}^{*}(v_1)\Psi_{\mu_2}^{*}(v_2)
=\sum_{\varepsilon_{\mu_1}+\varepsilon_{\mu_2}
=\varepsilon_{\mu_1'}+\varepsilon_{\mu_2'}}
W_{II}^*\left(\left.
\begin{array}{cc}
l+\bar{\varepsilon}_{\mu_1}+\bar{\varepsilon}_{\mu_2}&
l+\bar{\varepsilon}_{\mu_2'}\\
l+\bar{\varepsilon}_{\mu_1}&l
\end{array}\right|v_2-v_1 \right)
\Psi_{\mu_2'}^{*}(v_2)\Psi_{\mu_1'}^{*}(v_1), 
\label{bp1}
\end{eqnarray}
where 
$W_{II}^*\left(
\left.\begin{array}{cc}
a&b\\
c&d
\end{array}\right| u\right)$ are defined by
\begin{eqnarray*}
W_{II}^*
\left(\left.\begin{array}{cc}
a+2\bar{\varepsilon}_\mu & a+\bar{\varepsilon}_\mu\\
a+\bar{\varepsilon}_\mu & a
\end{array}\right| v \right)&=&S(v),\\
W_{II}^*
\left(\left.\begin{array}{cc}
a+\bar{\varepsilon}_\mu+\bar{\varepsilon}_\nu &
a+\bar{\varepsilon}_\mu\\
a+\bar{\varepsilon}_\nu & a
\end{array}\right| v\right)&=&
S(v)\frac{[v]'[a_{\mu \nu}-1]'}
{[v-1]'[a_{\mu \nu}]'},\\
W_{II}^*
\left(\left.\begin{array}{cc}
a+\bar{\varepsilon}_\mu+\bar{\varepsilon}_\nu &
a+\bar{\varepsilon}_\nu\\
a+\bar{\varepsilon}_\nu&a
\end{array}\right|v \right)&=&S(v)\frac{[v-a_{\mu \nu}]'[1]'}
{[v-1]'[a_{\mu \nu}]'}. 
\label{eq:BW2}
\end{eqnarray*}
%Here $S(v)$ and $\chi(v)$ are defined in (\ref{def:S})
%and (\ref{def:chi}).\\

Let us summarize the properties of $W^*_{II}$. 
Since it is obtaind from $W_{I}$ 
by replacing $r \rightarrow r-1$ up to a 
common scalar function, $W^*_{II}$ 
also satisfies the Yang-Baxter equation: 
\begin{equation}
\begin{array}{cc}
~ & 
\displaystyle \sum_{g} 
W^*_{II}\left( \left. 
\begin{array}{cc} d & e \\ c & g \end{array} \right| 
v_1 \right)
W^*_{II}\left( \left. 
\begin{array}{cc} c & g \\ b & a \end{array} \right| 
v_2 \right)
W^*_{II}\left( \left. 
\begin{array}{cc} e & f \\ g & a \end{array} \right| 
v_1 -v_2 \right) \\
~ & ~ \\
= & \displaystyle \sum_{g} 
W^*_{II}\left( \left. 
\begin{array}{cc} g & f \\ b & a \end{array} \right| 
v_1 \right)
W^*_{II}\left( \left. 
\begin{array}{cc} d & e \\ g & f \end{array} \right| 
v_2 \right)
W^*_{II}\left( \left. 
\begin{array}{cc} d & g \\ c & b \end{array} \right| 
v_1 -v_2 \right)
\end{array} \label{YBE2}
\end{equation}
The first and the second inversion relations 
are as follows \cite{JMO}: 
\begin{equation}
\sum_g W^*_{II}\left( \left. 
\begin{array}{cc} c & g \\ b & a \end{array} \right| 
-v \right)
W^*_{II}\left( \left. 
\begin{array}{cc} c & d \\ g & a \end{array} \right| 
v \right) =\delta_{bd}, 
\label{eq:1st}
\end{equation}
\begin{equation}
\sum_g G^*_g W^*_{II}\left( \left. 
\begin{array}{cc} g & b \\ d & c \end{array} \right| 
n-v \right)
W^*_{II}\left( \left. 
\begin{array}{cc} g & d \\ b & a \end{array} \right| 
v \right) =\delta_{ac}\frac{G^*_b G^*_d}{G^*_a}, 
\label{eq:2nd}
\end{equation}
where 
$$
G^*_a =\prod_{1\leqq \mu < \nu \leqq n} 
[a_{\mu\nu}]'. 
$$
The Boltzmann weights (\ref{eq:BW2}) have 
$\sigma$-invariant \cite{JMO} : 
\begin{equation}
W^*_{II}\left( \left. 
\begin{array}{cc} 
\sigma (c) & \sigma (d) \\ 
\sigma (b) & \sigma (a) \end{array} \right| 
v \right)=W^*_{II}\left( \left. 
\begin{array}{cc} c & d \\ b & a \end{array} \right| 
v \right)
\label{eq:s-inv}
\end{equation}
where $\sigma$ is the diagram automorphism of 
$A^{(1)}_{n-1}$ defined by $\sigma (\omega_\mu )
=\omega_{\mu +1}$. 

In section 3 we shall realize $\Psi_{\mu}^*(v)$ 
satisfying (\ref{bp2}) and (\ref{bp1}) 
in terms of free bosons. 

\subsection{Fused $A^{(1)}_{n-1}$ Boltzmann weight}

Let us introduce $m$-fold fused Boltzmann 
weights for $W^*_{II}$. See \cite{AJMP} 
concering the fused Boltzmann weights for $W_I$. 

Let $\Lambda =\{ \lambda_1 , \cdots , \lambda_m \}$ 
be a subset of $N=\{ 1, \cdots , n\}$ 
such that $\lambda_1 <\cdots <\lambda_m$. 
For $\kappa , \mu \in N$ 
set $\mu =\kappa$ if $\kappa \in \Lambda$, 
otherwise set $\mu \in \Lambda\cup \{\kappa \}$. 
For given $\kappa , \mu, \Lambda$ 
let $1\leqq \nu_1 <\cdots < \nu_m \leqq n$ 
be such that $\bar{\varepsilon}_\mu 
+\bar{\varepsilon}_{\nu_1}+ 
\cdots +\bar{\varepsilon}_{\nu_m}=
\bar{\varepsilon}_\kappa +
\bar{\varepsilon}_{\lambda_1}+ 
\cdots +\bar{\varepsilon}_{\lambda_m}$. 

The fusion of $W^*_{II}$ in the horizontal 
direction is constructed as follows. 
Let $a, b, c(=c_0) , c_1 , 
\cdots c_{m-1}, c_{m}(=d) \in P$ satisfy 
$$
c=b+\bar{\varepsilon}_\mu , \,\,
c_{j-1}-c_{j}=\bar{\varepsilon}_{\lambda_j}\,
(1\leqq j\leqq m), \,\,
d=a+\bar{\varepsilon}_\kappa . 
$$
Note that $b=a+\bar{\varepsilon}_{\nu_1}+ 
\cdots +\bar{\varepsilon}_{\nu_m}$ from 
the definition of $\nu_j$'s. Let $\sigma 
\in \mathfrak{S}_m$ be a permutation of 
$(1, \cdots , m)$, and set 
$$
b_0^\sigma =b,\,\, b^{\sigma}_j =b^{\sigma}_{j-1}-
\bar{\varepsilon}_{\nu_{\sigma (j)}}\,
(1\leqq j\leqq m),\,\, b_m^\sigma =a. 
$$
Then $m$-fold anti-symmetric fusion of $W_{II}^*$ 
in the horizontal direction is given as 
\begin{equation}
W_{II}^{(1,m)}\left( \left. 
\begin{array}{cc} c & d \\ b & a \end{array} \right| 
v \right)=\sum_{\sigma \in \mathfrak{S}_m} 
\mbox{sgn}\,\sigma \prod_{j=1}^m 
W_{II}^{*}\left( \left. 
\begin{array}{cc} c_{j-1} & c_j \\ 
b_{j-1}^\sigma & b_j^\sigma \end{array} \right| 
v+\frac{m+1}{2}-j \right). 
\end{equation}
Note that $W_{II}^{(1,m-1)}$ is anti-symmetric 
with respect to $(\lambda_1 , \cdots , \lambda_m )$. 

Next consider the fusion in the vertical direction. 
We use same $\kappa, \mu$, $\lambda_j$'s and 
$\nu_j$'s as before. Now we set 
$$
b=a+\bar{\varepsilon}_\mu , \,\,
d_{j-1}-d_{j}=\bar{\varepsilon}_{\lambda_j}\,
(1\leqq j\leqq m), \,\,
c=d+\bar{\varepsilon}_\kappa , 
$$
where $d_0 =d, d_m =a$. We have 
$c=b+\bar{\varepsilon}_{\nu_1}+ 
\cdots +\bar{\varepsilon}_{\nu_m}$. 
For $\sigma\in\mathfrak{S}_m$ set
$$
c_0^\sigma =c,\,\, c^{\sigma}_j =
c^{\sigma}_{j-1}-
\bar{\varepsilon}_{\nu_{\sigma (j)}}\,
(1\leqq j\leqq m),\,\, c_m^\sigma =b. 
$$
Then $m$-fold anti-symmetric fusion of $W_{II}^*$ 
in the vertical direction is given as 
\begin{equation}
W_{II}^{(m,1)}\left( \left. 
\begin{array}{cc} c & d \\ b & a \end{array} \right| 
v \right)=\sum_{\sigma \in \mathfrak{S}_m} 
\mbox{sgn}\,\sigma \prod_{j=1}^m 
W_{II}^{*}\left( \left. 
\begin{array}{cc} c_{j-1}^\sigma & d_{j-1} \\ 
c_{j}^\sigma & d_j \end{array} \right| 
v-\frac{m+1}{2}+j \right). 
\end{equation}
Note that $W_{II}^{(1,m-1)}$ is anti-symmetric 
with respect to $(\lambda_1 , \cdots , \lambda_m )$. 

We further introduce the fusion of $W_{II}^*$ 
in both horizontal and vertical directions. 
Let $\{ \kappa_j\}_{1\leqq j\leqq m}$ and 
$\{ \mu_j \}_{1\leqq j\leqq m}$ be subsets 
of $N$ such that $\sharp \{ \kappa_j\}
=\sharp \{ \mu_j \}=m$. 
Let $\{ \lambda_j \}_{1\leqq j\leqq m'}$ and 
$\{ \nu_j \}_{1\leqq j\leqq m'}$ be subsets 
of $N$ such that $\sharp \{ \lambda_j\}
=\sharp \{ \nu_j \}=m'$. 
Let $a, b, c, d\in P$ satisfy 
$$
d=a+\sum_{j=1}^m \bar{\varepsilon}_{\kappa_j}, \, 
c=d+\sum_{j=1}^{m'} \bar{\varepsilon}_{\lambda_j}, \, 
b=a+\sum_{j=1}^{m'} \bar{\varepsilon}_{\nu_j}, \, 
c=b+\sum_{j=1}^m \bar{\varepsilon}_{\mu_j}, 
$$
where 
$$
\sum_{j=1}^m \bar{\varepsilon}_{\kappa_j}+
\sum_{j=1}^{m'} \bar{\varepsilon}_{\lambda_j}=
\sum_{j=1}^m \bar{\varepsilon}_{\mu_j}+
\sum_{j=1}^{m'} \bar{\varepsilon}_{\nu_j}. 
$$
The $m\times m'$--fold fusion of $W^*_{II}$ 
is defiend as the antisymmetrized product of 
the $m'$-fold fusion of $W^*_{II}$ in the 
horizontal direction: 
\begin{equation}
W_{II}^{(m,m')}\left( \left. 
\begin{array}{cc} c & d \\ b & a \end{array} \right| 
v \right)=\sum_{\sigma \in \mathfrak{S}_m} 
\mbox{sgn}\,\sigma \prod_{j=1}^m 
W_{II}^{(1,m')}\left( \left. 
\begin{array}{cc} c_{j-1}^\sigma & d_{j-1} \\ 
c_{j}^\sigma & d_j \end{array} \right| 
v-\frac{m+1}{2}+j \right), 
\end{equation}
where 
$$
c_0^\sigma =c,\,\, c^{\sigma}_j =
c^{\sigma}_{j-1}-
\bar{\varepsilon}_{\mu_{\sigma (j)}}\,
(1\leqq j\leqq m),\,\, c_m^\sigma =b. 
$$
The $W_{II}^{(m,m')}$ can be also 
defiend as the antisymmetrized product of 
the $m$-fold fusion of $W^*_{II}$ in the 
vertical direction: 
\begin{equation}
W_{II}^{(m,m')}\left( \left. 
\begin{array}{cc} c & d \\ b & a \end{array} \right| 
v \right)=\sum_{\sigma \in \mathfrak{S}_{m'}} 
\mbox{sgn}\,\sigma \prod_{j=1}^{m'} 
W_{II}^{(m,1)}\left( \left. 
\begin{array}{cc} c_{j-1} & c_j \\ 
b_{j-1}^\sigma & b_j^\sigma \end{array} \right| 
v+\frac{m+1}{2}-j \right), 
\end{equation}
where 
$$
b_0^\sigma =b,\,\, b^{\sigma}_j =b^{\sigma}_{j-1}-
\bar{\varepsilon}_{\nu_{\sigma (j)}}\,
(1\leqq j\leqq m'),\,\, b_{m'}^\sigma =a. 
$$

\subsection{Fusion of the dual type II vertex operators} 

Here we introduce the fusion of the dual type II 
vertex operator $\Psi^*_\mu (v)$. Let 
$\Lambda =\{ \lambda_1 , \cdots , \lambda_m \}$ 
be the subset of $N=\{ 1, \cdots , n\}$ 
such that $\lambda_1 < \cdots < \lambda_m$. When 
$b=a+\bar{\varepsilon}_{\lambda_1} +\cdots +
\bar{\varepsilon}_{\lambda_m}$ we define the $m$-fold 
fused type II vertex operator $\Psi^{(m)}_\Lambda$ 
as follows: 
\begin{equation}
\Psi^{(m)}_\Lambda (v):=
\sum_{\sigma\in \mathfrak{S}_m} \mbox{sgn}\,
\sigma \Psi^*_{\lambda_{\sigma (1)}} 
\left( v+\frac{m-1}{2}\right) 
\Psi^*_{\lambda_{\sigma (2)}} 
\left( v+\frac{m-3}{2}\right) 
\cdots \Psi^*_{\lambda_{\sigma (m)}} 
\left( v-\frac{m-1}{2}\right). 
\label{df:Psi}
\end{equation}
It is clear from the definition 
that the following commutation relations hold 
\begin{equation}
\Psi^{(m)}_{\Lambda_1} (v)
\Psi^{(m')}_{\Lambda_2} (v' )
=\sum_d W_{II}^{(m,m')}\left( \left. 
\begin{array}{cc} c & d \\ b & a \end{array} \right| 
v' -v \right)
\Psi^{(m')}_{\Lambda'_2} (v' )
\Psi^{(m)}_{\Lambda'_1} (v), 
\label{eq:mm'-CR}
\end{equation}
if $\Psi^*_\mu$'s satisfy the commutation relations 
(\ref{bp2}). 

For the special case 
$\Lambda=\{ 1, \cdots , m\}$ and 
$\Lambda_\mu =\Lambda\backslash\{ \mu \}$ we define 
\begin{equation}
\Psi^{(m-1)}_\mu (v)=
\left( \prod_{j=1}^{m-1} c_j^{-1} \right) 
\Psi^{(m-1)}_{\Lambda_\mu}(v) ~~~~ 
(1\leqq \mu \leqq m).
\label{df:Psi-m}
\end{equation}
where
\begin{eqnarray}
c_j=
x^{-\frac{r}{r-1}\frac{j(1-j)}{2n}}
\frac{(x^{-2};x^{2r-2})_\infty^j
(x^{2r-2};x^{2r-2})_\infty^{2j-3}}{g_{j-1}^*(x^{-j})},
~~\mbox{for}~j=1,\cdots,n-1.
\label{df:c_j}
\end{eqnarray}
In section 3 we introduce the type II 
vertex operator $\bar{\Psi}^{(m-1)}_\mu$ 
in terms of free bosons. 
In section 4 we shall prove 
$\bar{\Psi}^{(m-1)}_\mu$ coincides with $\Psi^{(m-1)}_{\mu}$ 
up to a constant. 

\section{Bosonization of the 
type II vertex operators}

\subsection{Bosons}

Let us consider the bosons
$B_m^j\,(1\leqq j \leqq n-1, m \in \mathbb{Z}
\backslash \{0\})$
with the commutation relations
\begin{equation}
[B_m^j,B_{m'}^k]
=\left\{ \begin{array}{l} 
m\dfrac{[(n-1)m]_x}{[nm]_x}
\dfrac{[(r-1)m]_x}{[rm]_x}\delta_{m+m',0},~(j=k)\\
-mx^{{\rm sgn}(j-k)nm}\dfrac{[m]_x}{[nm]_x}
\dfrac{[(r-1)m]_x}{[rm]_x}\delta_{m+m',0},~(j\neq k), 
\end{array} \right. 
\label{eq:comm-B}
\end{equation}
where the symbol $[a]_x$ stands for
$(x^a-x^{-a})/(x-x^{-1})$.
Define $B_m^n$ by
\begin{eqnarray*}
\sum_{j=1}^n x^{-2jm}B_m^j=0.
\end{eqnarray*}
Then the commutation relations (\ref{eq:comm-B}) 
holds for all $1\leqq j,k \leqq n$.
These oscillators were introduced in \cite{FL,AKOS}. 

Define the dressed bosons 
$A_m^j\,(1\leqq j \leqq n, m \in \mathbb{Z}
\backslash\{0\})$ by
\begin{equation}
A_m^j=(-1)^m \frac{[rm]_x}{[(r-1)m]_x} B_m^j. 
\label{eq:df-A}
\end{equation}
The expression (\ref{eq:df-A}) 
for $n=2$ was already given in \cite{MW}. 
For $\alpha, \beta \in P$ let us define 
the zero mode operators $P_\alpha, Q_\beta$ 
with the commutation relations 
\begin{equation*}
[iP_{\alpha},Q_{\beta}]=\langle \alpha,\beta \rangle, 
~~~~ [P_{\alpha}, B_m^j ]=
[Q_{\beta}, B_m^j ]=0. 
\end{equation*}

~\\
We will deal with the bosonic Fock spaces 
${\cal{F}}_{l,k}, (l,k \in P)$
generated by $B_{-m}^j (m>0)$
over the vacuum vectors $|l,k\rangle$ :
\begin{eqnarray*}
{\cal{F}}_{l,k}=
\mathbb{C}[\{ B_{-1}^j, B_{-2}^j,\cdots \}_{1\leqq j \leqq n}]|l,k\rangle,
\end{eqnarray*}
where
\begin{eqnarray*}
B_m^j|k,l\rangle&=&0 ~(m>0),\\
P_{\alpha}|l,k\rangle &=&\langle \alpha,
\sqrt{\frac{r}{r-1}}l-\sqrt{\frac{r-1}{r}}k\rangle
|l,k\rangle,\\
|l,k\rangle&=&e^{i\sqrt{\frac{r}{r-1}}
Q_l-i\sqrt{\frac{r-1}{r}}Q_k}|0,0\rangle.
\end{eqnarray*}

\subsection{Basic Operator}

Let us define the basic operators for $j=1,\cdots,n-1$
\begin{eqnarray}
U_{-\alpha_j}(z)&=&
\exp\left(i\sqrt{\frac{r-1}{r}}(Q_{\alpha_j}
-iP_{\alpha_j}{\rm log} z)\right)
:\exp\left(\sum_{m \neq 0}\frac{1}{m}
(B_m^j-B_m^{j+1})(x^jz)^{-m}\right):, \\
U_{\omega_j}(z)&=&
\exp\left(-i\sqrt{\frac{r-1}{r}}(Q_{\omega_j}
-iP_{\omega_j}{\rm log}z)\right)
:\exp\left(
-\sum_{m\neq 0}\frac{1}{m}\sum_{k=1}^j
x^{(j-2k+1)m}B_m^k z^{-m}\right):, 
\end{eqnarray}
and
\begin{eqnarray}
V_{-\alpha_j}(z)&=&
\exp\left(-i\sqrt{\frac{r}{r-1}}(Q_{\alpha_j}
-iP_{\alpha_j}\log z)\right)
:\exp\left(
-\sum_{m\neq 0}\frac{1}{m}(A_m^j-A_m^{j+1})
x^{-mj}z^{-m}\right):, \\ 
V_{\omega_j}(z)&=&
\exp\left(
i\sqrt{\frac{r}{r-1}}(Q_{\omega_j}
-iP_{\omega_j}\log z)\right)
:\exp\left(
\sum_{m\neq 0}\frac{1}{m}
\sum_{k=1}^{j}x^{(j-2k+1)m}A_m^k z^{-m}
\right):.
\end{eqnarray}

Note that the operator $U_{-\alpha_j}(z)$ and 
$U_{\omega_j}(z) \,(j=1,\cdots, n-1)$
for $A_{n-1}^{(1)}$ face model
were introduced in \cite{AJMP}, and that 
the operator $V_{-\alpha_1}(z)$
and $V_{\omega_1}(z)$ 
for the $A_1^{(1)}$ face model were introduced 
in \cite{MW}. 

We will use the variable $v$ such that 
$z=x^{2v}$, and set 
\begin{equation}
\xi_j(v)=U_{-\alpha_j}(z),~~
\eta_j(v)=U_{\omega_j}(v), ~~
\xi_j^*(v)=V_{-\alpha_j}(z),~~
\eta_j^*(v)=V_{\omega_j}(z),
\end{equation}
for $j=1,\cdots,n-1$. These operators satisfy 
the following commutation relations: 
\begin{eqnarray}
\xi_j(v_1)\xi_j(v_2)&=&
h(v_1-v_2)\xi_j(v_2)\xi_j(v_1),\\
\xi_j^*(v_1)\xi_{j}^*(v_2)&=&
h^*(v_1-v_2)\xi_{j}^*(v_2)\xi_j^*(v_1), 
\label{eq:x*jj} \\
\xi_j(v_1)\xi_{j+1}(v_2)
&=&-f(v_1-v_2,0)\xi_{j+1}(v_2)\xi_j(v_1),\\
\xi_j^*(v_1)\xi_{j+1}^*(v_2)&=&
-f^*(v_1-v_2,0)\xi_{j+1}^*(v_2)\xi_j^*(v_1), 
\label{eq:x*jj+1} \\
\left[ \xi_j(v_1), \xi_{k}(v_2)\right] &=&
\left[ \xi_j^*(v_1), \xi_{k}^*(v_2)\right] =
0,~~(|j-k|>1), 
\label{eq:x*jk} \\
\xi_j(v_1)\eta_j(v_2)&=&
-f(v_1-v_2,0)\eta_j(v_2)\xi_j(v_1), \\
\xi_{j}^*(v_1)\eta_j^*(v_2)
&=&-f^*(v_1-v_2,0)\eta_j^*(v_2)\xi^*_j(v_1), 
\label{eq:h*x*} \\
\left[ \xi_j(v_1), \eta_{k}(v_2) \right] &=&
\left[ \xi^*_j(v_1), \eta^*_k(v_2) \right] =0,
~~(j\neq k), \label{eq:h*x*ne} \\
\left[ \xi_j(v_1), \xi_k^*(v_2) \right] &=&
\left[ \xi_j(v_1), \eta_k^*(v_2) \right] =
\left[ \eta_j(v_1), \xi_k^*(v_2) \right] =0,~~
(j,k=1,\cdots,n-1), \label{eq:hx*} \\
\eta_j(v_1)\eta_j(v_2)&=&r_j(v_1-v_2)
\eta_j(v_2)\eta_j(v_1), ~~ r_1 (v)=R(v), \\
\eta_j^*(v_1)\eta_j^*(v_2)&=&
s_j (v_2-v_1)\eta_j^*(v_2)\eta_j^*(v_1), ~~ 
s_1 (v)=S(v), \label{eq:h*11} \\
\eta_j(v_1)\eta_j^*(v_2)&=&
\chi_j (v_1-v_2)\eta_j^*(v_2)
\eta_j(v_1), ~~ \chi_1 (v)=\chi (v). \label{eq:hh*}
\end{eqnarray}

\subsection{Type I vertex operators}

In the sequel we set
\begin{eqnarray*}
\pi_\mu=\sqrt{r(r-1)}P_{\bar{\varepsilon}_\mu},~
\pi_{\mu \nu}=\pi_\mu-\pi_\nu.
\end{eqnarray*}
The $\pi_{\mu \nu}$
acts on ${\cal{F}}_{l,k}$ as an integer 
$\langle \varepsilon_\mu -\varepsilon_\nu,
rl-(r-1)k\rangle$.

For $1 \leqq \mu \leqq n$ 
define the type I vertex operator \cite{AJMP} by
\begin{eqnarray}
\Phi_\mu(v_0 )=\oint
\prod_{j=1}^{\mu-1}\frac{dz_j}{2\pi i z_j}
\eta_1(v_0 )\xi_1(v_1)\cdots \xi_{\mu-1}(v_{\mu-1})
\prod_{j=1}^{\mu-1}f(v_j-v_{j-1},\pi_{j \mu}), 
\label{eq:type-I}
\end{eqnarray}
where $z_j=x^{2v_j}$. 
The integral contour for $z_j$-integration 
encircles the poles at $z_j=x^{1+2kr}z_{j-1}\,(k 
\in \mathbb{Z}_{\geqq 0})$, but not the poles at 
$z_j=x^{-1-2kr}z_{j-1}\,(k \in 
\mathbb{Z}_{\geqq 0})$. 

Note that 
\begin{equation}
\Phi_\mu(v_0 ): {\cal{F}}_{l,k} 
\longrightarrow {\cal{F}}_{l,k+\bar{\varepsilon}_\mu}. 
\label{eq:k-shift}
\end{equation}
We thus denote the operator (\ref{eq:type-I}) by 
$\Phi_\mu^{(k+\bar{\varepsilon}_\mu,k)}(v_0 )$ 
on the bosonic Fock space ${\cal{F}}_{l,k}$. 
The commutation relations (\ref{eq:CR-I}) hold on 
${\cal{F}}_{l,k}$. 

For $1 \leqq \mu \leqq n$ define the dual 
type I vertex operator by
\begin{eqnarray}
\bar{\Phi}_\mu^{*(m-1)}(v_m )=\oint
\prod_{j=\mu}^{m-1}\frac{dz_j}{2\pi i z_j}
\eta_{m-1}(v_m )\xi_{m-1}(v_{m-1})
\cdots \xi_{\mu}(v_{\mu})
\prod_{j=\mu+1}^{m}f(v_{j-1}-v_{j},\pi_{\mu j}), 
\label{eq:type-I*}
\end{eqnarray}
where $z_j=x^{2v_j}$. 
The integral contour for $z_j$-integration 
encircles the poles at $z_j=x^{1+2kr}z_{j+1}\,
(k \in \mathbb{Z}_{\geqq 0})$, but not 
the poles at $z_j=x^{-1-2kr}z_{j+1}\,(k \in 
\mathbb{Z}_{\geqq 0})$. 

The operators (\ref{eq:type-I*}) is an 
operator such that 
\begin{equation}
\bar{\Phi}_\mu^{*(m-1)}(v_m ): 
{\cal F}_{l,k}\longrightarrow 
{\cal{F}}_{l,k+\bar{\varepsilon}_1 + 
\cdots +\bar{\varepsilon}_m 
-\bar{\varepsilon}_\mu}. 
\end{equation}
In particular $\bar{\Phi}_\mu^{*(n-1)}(v_n ): 
{\cal F}_{l,k}\longrightarrow 
{\cal{F}}_{l,k-\bar{\varepsilon}_\mu}$ 
for $m=n$. 

\subsection{Type II vertex operator}

In this subsection we introduce the type II vertex operator
for the $A_{n-1}^{(1)}$ face model.
For $1 \leqq \mu \leqq n$, define the dual type II vertex
operators by
\begin{equation}
\begin{array}{rcl}
\Psi_{\mu}^*(v_0 )&=&\displaystyle\oint
\prod_{j=1}^{\mu-1}\frac{dz_j}{2\pi i z_j}\eta_1^*(v_0 )
\xi_1^*(v_1)\cdots \xi_{\mu-1}^*(v_{\mu-1})
\prod_{j=1}^{\mu-1}f^*(v_j-v_{j-1},\pi_{j \mu}), \\
&=&
(-1)^{\mu-1} \displaystyle\oint \prod_{j=1}^{\mu-1}
\frac{dz_j}{2\pi i z_j}
\xi^*_{\mu-1}(v_{\mu-1})\cdots \xi_1^*(v_1)
\eta_1^*(v_0)\prod_{j=1}^{\mu-1}f^*(v_{j-1}-v_j,1-\pi_{j\mu}), 
\label{eq:II*}
\end{array}
\end{equation}
where $z_j=x^{2v_j}$. 
The second equality follows from 
(\ref{eq:x*jj+1}), (\ref{eq:x*jk}), (\ref{eq:h*x*}) 
and (\ref{eq:h*x*ne}). 
The integral contour for $z_j$-integration encloses 
the poles at $z_j=x^{-1+2k(r-1)}z_{j-1}\,
(k \in \mathbb{Z}_{\geqq 0})$, but not 
the poles at $z_j=x^{1-2k(r-1)}z_{j-1}\,(k \in 
\mathbb{Z}_{\geqq 0})$. Note that (\ref{eq:II*}) is 
an operator 
\begin{equation}
\Psi_{\mu}^*(v_0 ): {\cal F}_{l,k}\longrightarrow 
{\cal F}_{l+\bar{\varepsilon}_\mu ,k}, 
\end{equation}
so that we denote $\Psi_{\mu}^*(v_0 )$ by 
$\Psi_{\mu}^{*(l+\bar{\varepsilon}_\mu ,l)}(v_0 )$ 
on the bosonic Fock space ${\cal F}_{l,k}$. 

For $1 \leqq \mu \leqq n$, define the type II vertex
operators by
\begin{eqnarray*}
\bar{\Psi}_{\mu}^{(m-1)}(v_m )&=&\oint
\prod_{j=\mu}^{m-1}\frac{dz_j}{2\pi i z_j}
\eta_{m-1}^*(v_m )
\xi_{m-1}^*(v_{m-1})\cdots \xi_{\mu}^*(v_{\mu})
\prod_{j=\mu+1}^{m}f^*(v_{j-1}-v_{j},\pi_{\mu j}),\\
&=&
(-1)^{m-\mu}\oint
\prod_{j=\mu}^{m-1}\frac{dz_j}{2\pi i z_j}
\xi_{\mu}^*(v_\mu)\cdots \xi_{m-1}^*(v_{m-1})
\eta_{m-1}^*(v_m)
\prod_{j=\mu+1}^mf^*(v_j-v_{j-1},1-\pi_{\mu j}), 
\label{eq:type-II}
\end{eqnarray*}
where $z_j=x^{2v_j}$. The second equality again 
follows from 
(\ref{eq:x*jj+1}), (\ref{eq:x*jk}), (\ref{eq:h*x*}) 
and (\ref{eq:h*x*ne}). 
The integral contour for $z_j$-integration encloses 
the poles at $z_j=x^{-1+2k(r-1)}z_{j+1}\,(k 
\in \mathbb{Z}_{\geqq 0})$, but not 
the poles at $z_j=x^{1-2k(r-1)}z_{j+1}\,(k \in 
\mathbb{Z}_{\geqq 0}$). 
The $\bar{\Psi}_{\mu}^{(m-1)}(v_m )$ is an operator 
\begin{equation}
\bar{\Psi}_{\mu}^{(m-1)}(v_m ): 
{\cal F}_{l,k}\longrightarrow 
{\cal F}_{l+\bar{\varepsilon}_1 + 
\cdots +\bar{\varepsilon}_m 
-\bar{\varepsilon}_\mu ,k}. 
\end{equation}
In particular $\bar{\Psi}_\mu^{*(n-1)}(v_n ): 
{\cal F}_{l,k}\longrightarrow 
{\cal{F}}_{l-\bar{\varepsilon}_\mu ,k}$ 
for $m=n$. 

\subsection{Proof of commutation relations}

In this subsection we shall show that 
(\ref{eq:II*}) gives a bosonization of the 
dual type II vertex operators for the $A^{(1)}_{n-1}$ 
face model. For that purpose we prove the commutation 
relations (\ref{bp2}) and (\ref{bp1}). 

\begin{prop} Type I and Type II 
vertex operators commute modulo 
the `energy functions' $\chi_m$: 
\begin{eqnarray*}
\Phi_{\mu_1}(v_1)\Psi_{\mu_2}^{*}(v_2)
&=&\chi(v_1-v_2)
\Psi_{\mu_2}^{*}(v_2)\Phi_{\mu_1}(v_1),\\
\bar{\Phi}_{\mu_1}^{*(m-1)}(v_1)
\bar{\Psi}_{\mu_2}^{(m-1)}(v_2)
&=&\chi_{m-1}(v_1-v_2)
\bar{\Psi}_{\mu_2}^{(m-1)}(v_2)\bar{\Phi}_{\mu_1}^{*(m-1)}
(v_1).
\end{eqnarray*}
\end{prop}

{\sl Proof}. ~~ It is clear from 
(\ref{eq:hx*}) and (\ref{eq:hh*}). $\Box$

\begin{thm} The operators (\ref{eq:II*}) satisfy 
the commutation relations (\ref{bp1}). 
\label{thm:CR-II}
\end{thm}

{\sl Proof}. ~~ The claim of the theorem is 
equivalent to the following equations 
\begin{eqnarray}
\Psi_\mu^*(v_0)\Psi_\mu^*(v'_0)&=&
S(v'_0 -v_0 )\Psi_\mu^*(v'_0) 
\Psi_\mu^*(v_0), \label{eq:mumu} \\
\Psi_\mu^*(v_0)\Psi_\nu^*(v'_0)&=&
S(v'_0 -v_0 )\left\{ \Psi_\nu^*(v'_0) 
\Psi_\mu^*(v_0) 
b^* (v'_0 -v_0 , \pi_{\mu\nu}) \right. 
\nonumber \\
&+&
\left. \Psi_\mu^*(v'_0) 
\Psi_\nu^*(v_0) 
c^* (v'_0 -v_0 , \pi_{\mu\nu}) \right\} 
~~~~ (\mu \neq \nu ), \label{eq:munu}
\end{eqnarray}
where 
\begin{equation}
b^* (v, w)=\frac{[v]'[w-1]'}
{[v-1]'[w]'}, ~~~~
c^* (v, w)=\frac{[v-w]'[1]'}
{[v-1]'[w]'}. 
\end{equation}

Let us first prove (\ref{eq:mumu}). By using 
\begin{equation}
f^* (v, \pi_{j\mu}) \eta_1^*(v'_0)
\xi_1^*(v'_1)\cdots \xi_{\nu -1}^*(v'_{\nu -1})
=\eta_1^*(v'_0)
\xi_1^*(v'_1)\cdots \xi_{\nu -1}^*(v'_{\nu -1})
f^* (v, \pi_{j\mu}+r(\delta_{j\nu}-\delta_{\mu\nu})), 
\end{equation}
and the commutation relations 
(\ref{eq:x*jj+1}), (\ref{eq:x*jk}), 
(\ref{eq:h*x*}) we have 
\begin{equation}
\begin{array}{rcl}
\Psi_\mu^*(v_0) \Psi_\mu^*(v'_0) 
&=&
\displaystyle\oint \prod_{j=1}^{\mu -1} 
\dfrac{dz_j}{2\pi iz_j}\dfrac{dz'_j}{2\pi iz'_j} 
\eta^*_1 (v_0 )\eta^*_1 (v'_0 )
\xi^*_1 (v_1 )\xi^*_1 (v'_1 ) \cdots 
\xi^*_{\mu -1}(v_{\mu -1})
\xi^*_{\mu -1}(v'_{\mu -1}) \\
&\times &
\displaystyle\prod_{j=1}^{\mu -1} 
f^* (v_j -v'_{j-1}, 0)
f^* (v_j -v_{j-1}, \pi_{j\mu}-1)
f^* (v'_j -v'_{j-1}, \pi_{j\mu}). 
\end{array}
\label{eq:p-mumu}
\end{equation}
Since $v_1 , v'_1 , \cdots , v_{\mu -1}, 
v'_{\mu -1}$ are integral variables 
we can deform an integrand without 
changing the value of the integral. 
Actually the following 
deformation is allowed for any function 
$F(v_j , v'_j )$ coupled to 
$\xi^*_j (v_j )\xi^*_j (v'_j )$: 
\begin{equation}
\begin{array}{cl}
&\displaystyle\oint
\dfrac{dz_j}{2\pi iz_j}\dfrac{dz'_j}{2\pi iz'_j} 
\xi^*_j (v_j )\xi^*_j (v'_j )F(v_j , v'_j ) \\
=&
\displaystyle\oint
\dfrac{dz_j}{2\pi iz_j}\dfrac{dz'_j}{2\pi iz'_j} 
\xi^*_j (v'_j )\xi^*_j (v_j )
h^* (v_j -v'_j )F(v_j , v'_j ) \\
=& \dfrac{1}{2} 
\displaystyle\oint
\dfrac{dz_j}{2\pi iz_j}\dfrac{dz'_j}{2\pi iz'_j} 
\xi^*_j (v_j )\xi^*_j (v'_j )
\left\{ F(v_j , v'_j )+h^* (v'_j -v_j )F(v'_j , v_j )
\right\}, 
\end{array}
\label{eq:weak}
\end{equation}
where we use (\ref{eq:x*jj}). 
We thus denote $F(v_j , v'_j )\sim F'(v_j , v'_j )$ 
if $F(v_j , v'_j )$ and $F'(v_j , v'_j )$ satisfy 
$$
F(v_j , v'_j )+h^* (v'_j -v_j )F(v'_j , v_j )=
F'(v_j , v'_j )+h^* (v'_j -v_j )F'(v'_j , v_j ). 
$$
{}From (\ref{eq:p-mumu}), (\ref{eq:h*11}) and 
(\ref{eq:weak}) 
we can prove (\ref{eq:mumu}) by showing 
\begin{equation}
f^*_{11}(v_0 , v'_0, v_1 , v'_1)\sim 
f^*_{11}(v'_0 , v_0, v_1 , v'_1), 
\label{eq:f_11}
\end{equation}
where 
$$
f^*_{11}(v_0 , v'_0, v_1 , v'_1)=
f^* (v_1 -v'_{0}, 0)
f^* (v_1 -v_{0}, w-1)
f^* (v'_1 -v'_{0}, w). 
$$
Let 
$$
\tilde{f}^*_{11}(v_0 , v'_0, v_1 , v'_1)=
f^*_{11}(v_0 , v'_0, v_1 , v'_1)+
h^*(v'_1 -v_1)f^*_{11}(v_0 , v'_0, v'_1 , v_1)
$$ 
and set 
$$
F^*_{11}(v_1) =
\tilde{f}^*_{11}(v_0 , v'_0, v_1 , v'_1)-
\tilde{f}^*_{11}(v'_0 , v_0, v_1 , v'_1). 
$$
Since the residues at $v_1 =v_0 -\frac{1}{2}$, 
$v_1 =v'_0 -\frac{1}{2}$, 
$v_1 =v'_1 -1$ vanish, $F^*_{11}$ is a 
regular double periodic function of $v_1$, 
and hence a constant. We therefore get 
\begin{equation}
F^*_{11}(v_1 )=F^*_{11}(v'_0 +\tfrac{1}{2}-w)
=0, 
\label{eq:F_11}
\end{equation}
which implies (\ref{eq:f_11}). 

Next we prove (\ref{eq:munu}) for $\mu < 
\nu$. (We can show (\ref{eq:munu}) for $\mu >\nu$ 
in a similar manner.) 
When $\mu =1<\nu$, we have 
\begin{equation}
\begin{array}{rcl}
\Psi_1^*(v_0) \Psi_\nu^*(v'_0) 
&=&
\displaystyle\oint \prod_{j=1}^{\nu -1} 
\dfrac{dz'_j}{2\pi iz'_j} 
\eta^*_1 (v_0 )\eta^*_1 (v'_0 )
\xi^*_1 (v'_1 ) \cdots 
\xi^*_{\nu -1}(v'_{\nu -1}) 
\displaystyle\prod_{j=1}^{\nu -1} 
f^* (v'_j -v'_{j-1}, \pi_{j\nu}), \\
\Psi_\nu^*(v'_0) \Psi_1^*(v'_0) 
&=&
\displaystyle\oint \prod_{j=1}^{\nu -1} 
\dfrac{dz'_j}{2\pi iz'_j} 
\eta^*_1 (v'_0 )\eta^*_1 (v_0 )
\xi^*_1 (v'_1 ) \cdots 
\xi^*_{\nu -1}(v'_{\nu -1}) \\
&\times & f^* (v'_1 -v'_{0}, \pi_{1\nu}-1)
f^* (v'_1 -v_{0}, 0)
\displaystyle\prod_{j=2}^{\nu -1} 
f^* (v'_j -v'_{j-1}, \pi_{j\nu}). 
\end{array}
\label{eq:p-1nu}
\end{equation}
By the same argument as we show (\ref{eq:F_11}) 
we obtain 
\begin{equation}
\begin{array}{rcl}
f^* (v'_j -v'_{j-1}, \pi_{j\nu})&=&
b^*(v'_{j-1} -v_{j-1}, \pi_{j\nu})
f^* (v'_j -v'_{j-1}, \pi_{j\nu}+1)
f^* (v'_j -v_{j-1}, 0) \\
&+&
c^*(v'_{j-1} -v_{j-1}, \pi_{j\nu})
f^* (v'_j -v_{j-1}, \pi_{j\nu}). 
\end{array}
\label{eq:f-1nu}
\end{equation}
{}From (\ref{eq:p-1nu}) and (\ref{eq:f-1nu}) 
with $j=1$ we have (\ref{eq:munu}) for 
$\mu =1<\nu$. 

In order to prove (\ref{eq:munu}) for $1<\mu <\nu$ 
it is enough to show the following relation 
\begin{equation}
\begin{array}{cl}
& b^* (v'_0 -v_0 , \pi_{\mu\nu})
f^* (v'_1 -v_{0}, \pi_{1\mu})
f^* (v_1 -v_{0}, 0)
f^* (v_1 -v'_{0}, \pi_{1\nu}) 
f^* (v'_\mu -v'_{\mu -1}, 0) \\
\times & 
f^* (v'_\mu -v_{\mu -1}, \pi_{\mu\nu}+1 )
\displaystyle\prod_{j=2}^{\mu -1} 
f^* (v_j -v_{j-1}, \pi_{j\mu} )
f^* (v_j -v'_{j-1}, 0) 
f^* (v'_j -v'_{j-1}, \pi_{j\nu} ) \\
\sim &\displaystyle\prod_{j=1}^{\mu -1} 
f^* (v_j -v_{j-1}, \pi_{j\mu} )
\prod_{j=1}^{\mu -1} 
f^* (v_j -v'_{j-1}, 0) 
\prod_{j=1}^{\mu} 
f^* (v'_j -v'_{j-1}, \pi_{j\nu} ) \\
-& c^* (v'_0 -v_0 , \pi_{\mu\nu})
f^* (v_1 -v'_{0}, \pi_{1\mu})
f^* (v_1 -v_{0}, 0)
f^* (v'_1 -v_{0}, \pi_{1\nu}) \\
\times & \displaystyle\prod_{j=2}^{\mu -1} 
f^* (v_j -v_{j-1}, \pi_{j\mu} )
\prod_{j=2}^{\mu -1} 
f^* (v_j -v'_{j-1}, 0) 
\prod_{j=2}^{\mu} 
f^* (v'_j -v'_{j-1}, \pi_{j\nu} ). 
\end{array}
\label{eq:mu}
\end{equation}

We would like to prove (\ref{eq:mu}) 
by induction with respect to 
$\mu$. Set $\mu =2<\nu$. Then (\ref{eq:mu}) 
reduces to 
\begin{equation}
\begin{array}{cl}
&b^* (v'_0 -v_0 , \pi_{2\nu})
f^* (v'_2 -v'_{1}, \pi_{2\nu}+1 )
f^* (v'_2 -v_{1}, 0) \\
\times & f^* (v_1 -v_{0}, \pi_{12})
f^* (v'_1 -v_{0}, 0)
f^* (v'_1 -v'_{0}, \pi_{1\nu}) \\
\sim & f^* (v'_2 -v'_{1}, \pi_{2\nu} )
f^* (v_1 -v_{0}, \pi_{12} )
f^* (v_1 -v'_{0}, 0) 
f^* (v'_1 -v'_{0}, \pi_{1\nu} ) \\
-& c^* (v'_0 -v_0 , \pi_{2\nu})
f^* (v'_2 -v'_{1}, \pi_{2\nu} )
f^* (v_1 -v'_{0}, \pi_{12})
f^* (v_1 -v_{0}, 0)
f^* (v'_1 -v_{0}, \pi_{1\nu}). 
\end{array}
\label{eq:2nu}
\end{equation}
Here we exchange $v_1$ and $v'_1$ in the term 
including $b^*$ because they are integral variables. 
Owing to (\ref{eq:f-1nu}) with $j=2$, 
(\ref{eq:2nu}) is equivalent to 
\begin{equation}
\begin{array}{cl}
&f^* (v'_2 -v'_{1}, \pi_{2\nu})
f^* (v'_1 -v'_{0}, \pi_{1\nu} ) 
f^* (v_1 -v_{0}, \pi_{12} ) \\ 
\times & \left\{ 
\dfrac{b^* (v'_0 -v_0 , \pi_{2\nu})}{
b^* (v_1 -v'_1 , \pi_{2\nu})}
b^* (v_1 -v'_1 , \pi_{2\nu})
f^* (v_1 -v'_{0}, 0) 
-f^* (v_1 -v'_{0}, 0) \right\} \\
+ & f^* (v'_2 -v'_{1}, \pi_{2\nu} ) 
f^* (v_1 -v_{0}, 0)
\left\{ c^* (v'_0 -v_0 , \pi_{2\nu})
f^* (v_1 -v'_{0}, \pi_{12})
f^* (v'_1 -v_{0}, \pi_{1\nu}) \right. \\
- & \left. \dfrac{c^* (v_1 -v'_1 , \pi_{2\nu})}
{b^* (v_1 -v'_1 , \pi_{2\nu})}
b^* (v'_0 -v_0 , \pi_{2\nu})
f^* (v'_1 -v_{0}, \pi_{12})
f^* (v_1 -v'_{0}, \pi_{1\nu}) \right\} \\
\sim & 0, 
\end{array}
\label{eq:ch-2nu}
\end{equation}
where we use the relation 
$b^* (v'_1 -v_1 , w) \sim 
b^* (v_1 -v'_1 , w)$. 
By using the identities 
\begin{equation}
\dfrac{b^* (v'_0 -v_0 , w)}{
b^* (v_1 -v'_1 , w)}
\dfrac{f^* (v'_1 -v_0 , 0)}{
f^* (v_1 -v'_1 , 0)}-1=
\dfrac{[1]'[v_0 -v'_0 +v_1 -v'_1 ]'
[v_0 -v_1 +\frac{1}{2}]'[v'_0 -v'_1 -\frac{1}{2}]'}{
[v'_0 -v_0 -1]'[v_1 -v'_1 ]'
[v'_1 -v_0 +\frac{1}{2}]'[v_1 -v'_0 -\frac{1}{2}]'}, 
\label{eq:b-id}
\end{equation}
\begin{equation}
\begin{array}{cl}
&c^* (v'_0 -v_0 , w)-c^* (v_1 -v'_1 , w)
\dfrac{b^* (v'_0 -v_0 , w)}{
b^* (v_1 -v'_1 , w)}
\dfrac{f^* (v'_1 -v_0 , w+w')}{
f^* (v_1 -v'_0 , w+w')}
\dfrac{f^* (v'_1 -v_0 , w')}{
f^* (v_1 -v'_0 , w')} \\
=&\dfrac{[1]'[v_0 -v'_0 +v_1 -v'_1 ]'
[v_0 -v_1 +\frac{1}{2}-w']'[v'_1 -v'_0 -\frac{1}{2}
+w+w']'}{[v'_0 -v_0 -1]'[v_1 -v'_1 ]'
[v'_1 -v_0 -\frac{1}{2}+w+w']'[v_1 -v'_0 -\frac{1}{2}+w']'}, 
\end{array}
\label{eq:bc-id}
\end{equation}
we have (\ref{eq:ch-2nu}), which implies (\ref{eq:2nu}). 

Suppose $2<\mu <\nu$. From 
the assumption of the induction 
\begin{equation}
\begin{array}{cl}
& b^* (v_1 -v'_1 , \pi_{\mu\nu})
f^* (v'_\mu -v'_{\mu -1}, 0) 
f^* (v'_\mu -v_{\mu -1}, \pi_{\mu\nu}+1 ) \\
\times & 
\displaystyle\prod_{j=2}^{\mu -1} 
f^* (v_j -v_{j-1}, \pi_{j\mu} )
f^* (v_j -v'_{j-1}, 0) 
f^* (v'_j -v'_{j-1}, \pi_{j\nu} ) \\
\sim &f^* (v_2 -v'_{1}, \pi_{2\mu})
f^* (v_2 -v_{1}, 0)
f^* (v'_2 -v_{1}, \pi_{2\nu}) \\
\times &\displaystyle\prod_{j=3}^{\mu -1} 
f^* (v_j -v_{j-1}, \pi_{j\mu} )
\prod_{j=3}^{\mu -1} 
f^* (v_j -v'_{j-1}, 0) 
\prod_{j=3}^{\mu} 
f^* (v'_j -v'_{j-1}, \pi_{j\nu} ) \\
-& c^* (v_1 -v'_1 , \pi_{\mu\nu})
\displaystyle\prod_{j=2}^{\mu -1} 
f^* (v_j -v_{j-1}, \pi_{j\mu} )
\prod_{j=2}^{\mu -1} 
f^* (v_j -v'_{j-1}, 0) 
\prod_{j=2}^{\mu} 
f^* (v'_j -v'_{j-1}, \pi_{j\nu} ), 
\end{array}
\label{eq:mu-1}
\end{equation}
we have 
\begin{equation}
\begin{array}{rcl}
\mbox{LHS of $(\ref{eq:mu})$} & 
\sim & 
\dfrac{b^* (v'_0 -v_0 , \pi_{\mu\nu})}{
b^* (v_1 -v'_1 , \pi_{\mu\nu})}
\left\{ 
\displaystyle\prod_{j=1}^{\mu -1} 
f^* (v_j -v_{j-1}, \pi_{j\mu} )
\prod_{j=2}^{\mu -1} 
f^* (v_j -v'_{j-1}, 0) 
\prod_{j=1}^{\mu} 
f^* (v'_j -v'_{j-1}, \pi_{j\nu} ) \right. \\
&-& c^* (v_1 -v'_1 , \pi_{\mu\nu})
f^* (v'_1 -v_{0}, \pi_{1\mu})
f^* (v_1 -v_{0}, 0)
f^* (v_1 -v'_{0}, \pi_{1\nu}) \\
&\times & \left. 
\displaystyle\prod_{j=2}^{\mu -1} 
f^* (v_j -v_{j-1}, \pi_{j\mu} )
\prod_{j=2}^{\mu -1} 
f^* (v_j -v'_{j-1}, 0) 
\prod_{j=2}^{\mu} 
f^* (v'_j -v'_{j-1}, \pi_{j\nu} ) \right\}. 
\end{array}
\label{eq:ch-mu}
\end{equation}
Repeating the same procedure and using 
(\ref{eq:b-id},\ref{eq:bc-id}) as we show 
(\ref{eq:2nu}), we obtain (\ref{eq:mu}) 
for $\mu <\nu$. $\Box$

\section{Inversion and duality}

In this section we prove various properties 
of the vertex operators for the $A^{(1)}_{n-1}$ 
face model. Besides the formulae listed in the 
last section we will use the follwing formuale 
of normal ordering and commutation relations among the basic 
operators: 

\begin{eqnarray}
\eta_j^*(v_1)\xi_j^*(v_2)
&=&:\eta_j^*(v_1)\xi_j^*(v_2):
z_1^{-\frac{r}{r-1}}
\frac{(x^{2r-1}\frac{z_2}{z_1};x^{2r-2})_\infty}
{(x^{-1}\frac{z_2}{z_1};x^{2r-2})_\infty}, 
\label{eq:OPE-h*x*} \\
\xi_j^*(v_1)\eta_j^*(v_2)
&=&:\xi_j^*(v_1)\eta_j^*(v_2):
z_1^{-\frac{r}{r-1}}
\frac{(x^{2r-1}\frac{z_2}{z_1};x^{2r-2})_\infty}
{(x^{-1}\frac{z_2}{z_1};x^{2r-2})_\infty},\\
\xi^*_j(v_1)\eta^*_k(v_2)&=&:\eta^*_k(v_2)
\xi^*_j(v_1):
,~(j\neq k),\\
\eta^*_j(v_1)\xi^*_k(v_2)&=&:\xi^*_k(v_2)
\eta^*_j(v_1):
,~(j\neq k),\\
\xi_j^*(v_1)\xi_{j+1}^*(v_2)&=&
:\xi_{j+1}^*(v_2)\xi_j^*(v_1):
z_1^{-\frac{r}{r-1}}
\frac{(x^{2r-1}\frac{z_2}{z_1};x^{2r-2})_\infty}
{(x^{-1}\frac{z_2}{z_1};x^{2r-2})_\infty},\\
\xi_{j+1}^*(v_1)\xi_{j}^*(v_2)&=&
:\xi_{j}^*(v_1)\xi_{j+1}^*(v_2):
z_1^{-\frac{r}{r-1}}
\frac{(x^{2r-1}\frac{z_2}{z_1};x^{2r-2})_\infty}
{(x^{-1}\frac{z_2}{z_1};x^{2r-2})_\infty},\\
\xi_j^*(v_1)\xi_{k}^*(v_2)&=&
:\xi_{k}^*(v_2)\xi_j^*(v_1):,~~(|j-k|>1), 
\\
\eta_1^*(v_1)\eta_{l}^*(v_2)&=&
:\eta_{l}^*(v_2)\eta_1^*(v_1):
z_1^{\frac{r}{r-1}\frac{n-l}{n}}
g_l^*(z_2/z_1),\\
\eta_{l}^*(v_2)\eta_{1}^*(v_1)&=&
:\eta_{l}^*(v_2)\eta_1^*(v_1):
z_2^{\frac{r}{r-1}\frac{n-l}{n}}
g_l^*(z_1/z_2),\\
\eta_1^*(v_1)\eta_{l}^*(v_2)&=&
r_{l}^{*}(v_1-v_2)\eta_{l}^*(v_2)\eta_1^*(v_1),
\label{eq:h*1l}
\end{eqnarray}
\begin{eqnarray*}
\eta_j(v_1)\eta_j(v_2)
&=&z_1^{\frac{r-1}{r}\frac{j(n-j)}{n}}
\frac{\left\{x^{2r+2j}\frac{z_2}{z_1}\right\}_\infty
\left\{x^{2n+2r-2j}\frac{z_2}{z_1}\right\}_\infty
\left\{x^{2}\frac{z_2}{z_1}\right\}_\infty
\left\{x^{2n+2}\frac{z_2}{z_1}\right\}_\infty}
{\left\{x^{2j+2}\frac{z_2}{z_1}\right\}_\infty
\left\{x^{2n-2j+2}\frac{z_2}{z_1}\right\}_\infty
\left\{x^{2r}\frac{z_2}{z_1}\right\}_\infty
\left\{x^{2n+2r}\frac{z_2}{z_1}\right\}_\infty
}
:\eta_j(v_1)\eta_j(v_2):,\\
\eta_j^*(v_1)\eta_j^*(v_2)
&=&z_1^{\frac{r}{r-1}\frac{j(n-j)}{n}}
\frac{\left\{x^{2r+2j}\frac{z_2}{z_1}\right\}_\infty'
\left\{x^{2n+2r-2j}\frac{z_2}{z_1}\right\}_\infty'
\left\{\frac{z_2}{z_1}\right\}_\infty'
\left\{x^{2n}\frac{z_2}{z_1}\right\}_\infty'}
{\left\{x^{2j}\frac{z_2}{z_1}\right\}_\infty'
\left\{x^{2n-2j}\frac{z_2}{z_1}\right\}_\infty'
\left\{x^{2r}\frac{z_2}{z_1}\right\}_\infty'
\left\{x^{2n+2r}\frac{z_2}{z_1}\right\}_\infty'
}
:\eta_j^*(v_1)\eta_j^*(v_2):,\\
\eta_j(v_1)\eta_j^*(v_2)
&=&z_1^{\frac{j(j-n)}{n}}
\frac{(-x^{2j+1}\frac{z_2}{z_1};x^{2n},x^2)_\infty
(-x^{2n-2j+1}\frac{z_2}{z_1};x^{2n},x^2)_\infty
}{(-x\frac{z_2}{z_1};x^{2n},x^2)_\infty
(-x^{2n+1}\frac{z_2}{z_1};x^{2n},x^2)_\infty
}
:\eta_j(v_1)\eta_j^*(v_2):,\\
\eta_j^*(v_2)\eta_j(v_1)
&=&z_2^{\frac{j(j-n)}{n}}
\frac{(-x^{2j+1}\frac{z_1}{z_2};x^{2n},x^2)_\infty
(-x^{2n-2j+1}\frac{z_1}{z_2};x^{2n},x^2)_\infty
}{(-x\frac{z_1}{z_2};x^{2n},x^2)_\infty
(-x^{2n+1}\frac{z_1}{z_2};x^{2n},x^2)_\infty
}
:\eta_j(v_1)\eta_j^*(v_2):,
\end{eqnarray*}
where 
\begin{eqnarray*}
\left\{z\right\}_\infty
=(z;x^{2n},x^{2r},x^2)_\infty,~
\left\{z\right\}_\infty'
=(z;x^{2n},x^{2r-2},x^2)_\infty.
\end{eqnarray*}
Do not confuse $\{z\}$, $\{z\}'$ defined in 
(\ref{eq:{z}}) and 
$\{z\}_\infty$, $\{z\}'_\infty$, respectively. 

\subsection{Inversion relations} 

In this subsection we prove the following 
two theorems: 

\begin{thm}[Inversion identity] As 
$v' \to v -\frac{n}{2}$, 
\begin{eqnarray}
&&\sum_{\mu=1}^n
\bar{\Psi}_{\mu}^{(n-1)}(v)
\Psi_\mu^*(v')\prod_{j=1 \atop{j\neq \mu}}^n
{[\pi_{j\mu}]'}^{-1}=\frac{g_n'}{1-(x^n z')/z}, 
%+\mbox{regular terms at $v'=v-\frac{n}{2}$},
\label{bp6}\\
&&\sum_{\mu=1}^n
{\Psi}_{\mu}^{*}(v)
\bar{\Psi}_\mu^{(n-1)}(v')\prod_{j=1 \atop{j\neq \mu}}^n
{[\pi_{j\mu}]'}^{-1}=\frac{g_n'}{1-(x^n z')/z},
%+\mbox{regular terms at $v'=v-\frac{n}{2}$}, 
\label{bp7}
\end{eqnarray}
where 
\begin{eqnarray}
g_n'=(-1)^{n-1}
\frac{x^{\frac{r}{r-1}\frac{n(1-n)}{2n}}}
{(x^{2r-2};x^{2r-2})_\infty^{2n-3}
(x^{-2};x^{2r-2})_\infty^n}
\frac{(x^{2r};x^{2n},x^{2r-2})_\infty
(x^{-2};x^{2n},x^{2r-2})_\infty }
{(x^{2r-2};x^{2n},x^{2r-2})_\infty^2
(x^{2n},x^{2n})_\infty 
}
\label{def:g'}
\end{eqnarray}
\label{thm:inv-II}
\end{thm}

\begin{thm} As $v'\to v+\frac{n}{2}$, 
the product of vertex operators behaves like
\begin{eqnarray}
\bar{\Psi}_\mu^{(n-1)}(v)\Psi_\nu^*(v')
=\delta_{\mu\nu}\frac{g_n}{1-z'/(x^n z)}
\prod_{j=1}^n
\frac{[1-\pi_{j\mu}]'}{[1]'}+ 
(\mbox{regular terms at $v'=v+\frac{n}{2}$}), 
\label{bp5}
\end{eqnarray}
where 
\begin{eqnarray}
g_n=(-1)^{n-1}x^{\frac{r}{2(r-1)}(n-1)}
\left(\frac{(x^{2r};x^{2r-2})_\infty}
{(x^{2r-2};x^{2r-2})_\infty}\right)^n
\frac{(x^{2n+2r};x^{2n};x^{2r-2})_\infty 
(x^{2n-2};x^{2n}.x^{2r-2})_\infty
}
{(x^{2n};x^{2n},x^{2r-2})_\infty 
(x^{2n-2r-2};x^{2n},x^{2r-2})_\infty}. 
\label{def:g}
\end{eqnarray}
\label{thm:OPE-II}
\end{thm}

Let us begin from the following Lemma: 

\begin{lem} For $1\leqq \mu \leqq m$ we have
\begin{equation}
\begin{array}{cl}
&\bar{\Psi}_\mu^{(m-1)}(v_m )
\Psi^*_{\mu}(v'_0 ) \\
=&-r_{m-1}^*(v'-v)^{-1}
\displaystyle\sum_{\nu=1}^{m}
\Psi_\nu^*(v')\bar{\Psi}_{\nu}^{(m-1)}(v)
\frac{[v-v'-\frac{m}{2}-1-\pi_{\mu \nu}]'}
{[v-v'-\frac{m}{2}]'}
\prod_{j=1 \atop{j \neq \nu}}^m
\frac{[1-\pi_{j \mu}]'}{[\pi_{j \nu}]'}
\end{array}
\label{eqn:c11}
\end{equation}
\end{lem}

{\sl Proof}. ~~ From the commutation relations 
(\ref{eq:x*jj+1}), (\ref{eq:x*jk}), 
(\ref{eq:h*x*}), (\ref{eq:h*x*ne}) and 
(\ref{eq:h*1l}) 
\begin{equation}
\begin{array}{rcl}
\Psi_\mu^{(m-1)}(v_m) \Psi_\mu^*(v_0) 
&=&(-1)^{m-1}
\displaystyle\oint \prod_{j=1}^{m-1} 
\dfrac{dz_j}{2\pi iz_j} 
\eta^*_{m-1} (v_{m-1})
\xi^*_{m-1}(v_{m-1}) \cdots 
\xi^*_{1}(v_{1})
\eta^*_{1}(v_{0}) \\
&\times &
\displaystyle\prod_{j=1}^{m} 
f^* (v_{j-1} -v_{j}, 1-\pi_{j\mu}), 
\end{array}
\label{eq:mumu*}
\end{equation}
\begin{equation}
\begin{array}{rcl}
\Psi_\nu^{*}(v_0 ) \Psi_\mu^{(m-1)}(v_m ) 
&=& -
\displaystyle\oint \prod_{j=1}^{m-1} 
\dfrac{dz_j}{2\pi iz_j} 
\eta^*_{m-1} (v_{m-1})
\xi^*_{m-1}(v_{m-1}) \cdots 
\xi^*_{1}(v_{1})
\eta^*_{1}(v_{0}) \\
&\times & r^*_{m-1} (v_0 -v_m )
\displaystyle\prod_{j=1}^{m} 
f^* (v_{j-1} -v_{j}, \pi_{\nu j}). 
\end{array}
\label{eq:mu*mu}
\end{equation}
In order to show (\ref{eqn:c11}) let us 
consider the following elliptic function
\begin{eqnarray*}
F^*(u)=\frac{[v_m-v_0-\frac{m}{2}-1-\pi_\mu+u]'}
{[1+\pi_\mu-u]'}
\prod_{j=1}^m\frac{[v_{j-1}-v_j-\frac{1}{2}-\pi_j+u]'}
{[\pi_j-u]'}.
\end{eqnarray*}
Set the sum of all residues of $F^*$ 
in the period to be zero. Then we obtain 
\begin{equation}
\begin{array}{cl}
&[v_m-v_0-\frac{m}{2}]'
\displaystyle\prod_{j=1}^m
\frac{[v_{j-1}-v_j+\frac{1}{2}-\pi_j+\pi_\mu]'}
{[\pi_j-\pi_\mu-1]'}\\
=& -\displaystyle\sum_{\nu=1}^m 
\frac{[v_{\nu-1}-v_\nu-\frac{1}{2}]'
[v_m-v_0-\frac{m}{2}-1-\pi_\mu+\pi_\nu]'}
{[1+\pi_\mu-\pi_\nu]'}
\prod_{j=1 \atop{j\neq \nu}}^m
\frac{[v_{j-1}-v_j-\frac{1}{2}-\pi_j+\pi_\nu]'}
{[\pi_j-\pi_\nu]'}, 
\end{array}
\label{eq:lem1}
\end{equation}
which implies (\ref{eqn:c11}) because of 
(\ref{eq:mumu*}) and (\ref{eq:mu*mu}). $\Box$

\begin{lem} For $1\leqq m\leq n-1$
\begin{equation}
\begin{array}{cl}
&\displaystyle\eta_{m-1}^*\left(v+\frac{1}{2}\right)
\xi_{m-1}^*\left(v\right)\cdots
\xi_{1}^*\left(v-\frac{m-2}{2}\right)
\eta_{1}^*\left(v-\frac{m-1}{2}\right) \\
=&\displaystyle
x^{\frac{r}{r-1}\frac{m(1-m)}{2n}}
\left(\frac{(x^{2r-2};x^{2r-2})_\infty}
{(x^{-2};x^{2r-2})_\infty}\right)^m
g_{m-1}^*(x^{-m})\times \eta^*_m(v). 
\end{array}
\label{eq:p-bo<n}
\end{equation}
For $m=n$, as $v_n \to v+\frac{1}{2},~
v_{n-1}\to v, \cdots, v_0 \to v-\frac{n-1}{2}$,
\begin{equation}
\begin{array}{cl}
&\eta_{n-1}^*(v_n)
\xi_{n-1}^*\left(v_{n-1}\right)\cdots
\xi_{1}^*\left(v_1\right)
\eta_{1}^*\left(v_0\right) \\
\to&\displaystyle
x^{\frac{r}{r-1}\frac{n(1-n)}{2n}}
\left(\frac{(x^{2r-2};x^{2r-2})_\infty}
{(x^{-2};x^{2r-2})_\infty}\right)^n
g_{n-1}^*\left(\frac{z_0}{z_n}\right)\times \eta^*_n(v).
\end{array}
\label{eq:p-bo=n}
\end{equation}
\label{lem:prod}
\end{lem}

{\sl Proof}. The claim for $1\leqq m \leqq n-1$ 
follows from (\ref{eq:OPE-h*x*}--\ref{eq:h*1l}). 
For $m=n$ note that $g^*_{n-1}(z)$ has a pole at 
$z=x^{-n}$. We obtain (\ref{eq:p-bo=n}) by 
taking the limit $v_j =v-\frac{n-1-j}{2}\,
(0\leqq j\leqq n)$. $\Box$ 

\begin{lem} For $1\leqq m\leqq n-1$, the following 
relations hold: 
\begin{equation}
\begin{array}{cl}
&\displaystyle\sum_{\mu=1}^{m}
\bar{\Psi}_{\mu}^{(m-1)}\left(v+\frac{1}{2}\right)
\Psi_{\mu}^*\left(v-\frac{m-1}{2}\right)
\prod_{j=1 \atop{j\neq \mu}}^m{[\pi_{j \mu}]'}^{-1}
=c_m^{-1}\bar{\Psi}_{m+1}^{(m)}(v),
%\\
%=& \displaystyle
%\frac{x^{\frac{r}{r-1}\frac{m(1-m)}{2n}}
%g_{m-1}^*(x^{-m})}
%{(x^{-2};x^{2r-2})_\infty^m 
%(x^{2r-2};x^{2r-2})_\infty^{2m-3}}
%\, \eta^*_m(v) \\
%=& \displaystyle
%\frac{x^{\frac{r}{r-1}\frac{m(1-m)}{2n}}
%g_{m-1}^*(x^{-m})}
%{(x^{-2};x^{2r-2})_\infty^m 
%(x^{2r-2};x^{2r-2})_\infty^{2m-3}}
%\, \bar{\Psi}^{(m)}_{m+1} (v). 
\end{array}
\label{eq:m-inv-le}
\end{equation}
where $c_m$ is defined in (\ref{df:c_j}).
For $m=n$, we have
\begin{eqnarray}
\sum_{\mu=1}^{n}
\bar{\Psi}_{\mu}^{(n-1)}(v)
\Psi_{\mu}^*(v')
\prod_{j=1 \atop{j\neq \mu}}^n{[\pi_{j \mu}]'}^{-1}
=\frac{g_n'}{1-(x^nz')/z},
%+(\mbox{regular terms 
%at $v'=v-\frac{n}{2}$}), 
\label{eq:m-inv-n}
\end{eqnarray}
where $g_n'$ is defined in (\ref{def:g'}).
\label{prop:m-inv}
\end{lem}

{\sl Proof}. ~~ Let $v_0=v-\frac{m-1}{2} $ and 
$ v_m=v+\frac{1}{2} $. From (\ref{eq:mumu*}) 
we have 
\begin{equation}
\begin{array}{cl}
&\displaystyle\sum_{\mu=1}^{m} 
\bar{\Psi}_{\mu}^{(m-1)}(v+\frac{1}{2}) 
\Psi_{\mu}^*(v-\frac{m-1}{2}) 
\prod_{1\leqq j \leqq m \atop{j \neq \mu}} 
[\pi_{j \mu}]'{}^{-1} \\
=& \displaystyle\sum_{\mu=1}^{m}(-1)^{m-1}
\oint_{C_{\mu}} \prod_{j=1}^{m-1} \frac{dz_j}{2\pi i z_j}
\eta_{m-1}^*(v_m)
\xi^*_{m-1}(v_{m-1})\cdots \xi_1^*(v_1)
\eta_1^*(v_0) 
F_{\mu}^*(v_1,\cdots,v_{m-1}),
\label{eqn: c17}
\end{array}
\end{equation}
where 
\[ F_{\mu}^*(v_1, \cdots, v_{m-1}) = 
\prod_{1\leqq j \leqq m \atop{j \neq \mu}}
\frac{f^*(v_{j-1} - v_j , 1-\pi_{j \mu})}
{[\pi_{j \mu}]'} . \]
The contour $C_\mu$ is chosen as 
\begin{equation}
\begin{array}{ll}
+ \{ |z_j| = x^{-m+j+1}(|z|-j \varepsilon) \} \\
 + \{ |z_j-x^{-m+j-1}z|=\varepsilon \} - 
 \{ |z_j-x^{-m+j+3}z|=\varepsilon \} & 
 ( 1 \leqq j \leqq \mu -1 )\\
+ \{ |z_j| = x^{-m+j+1}(|z|+(m-j) \varepsilon) \} \\
 + \{ |z_j-x^{-m+j-1}z|=\varepsilon \} - 
 \{ |z_j-x^{-m+j+3}z|=\varepsilon \} 
 & ( \mu \leqq j \leqq m-1 )
\end{array}
\label{eqn: c18},
\end{equation}
for a small number $ \varepsilon > 0 $. 
Here the signs of the integral paths
represent the directions. The plus sign 
refers to an anti-clockwise contour, and 
the minus sign refers to a clockwise contour.

Set the sum of all residues in the period we have 
to be zero. Then we have
\begin{equation}
\sum_{\mu=1}^{m} F_{\mu}^* (v_1,\cdots,v_{m-1}) = 0 .
\label{eqn: c19}
\end{equation}
Note that the RHS of (\ref{eqn: c17}) vanishes 
from (\ref{eqn: c19}) 
if the exchange of the order of the sum and the 
integral is permited. 
In the neighbourhood of the contour $C_{\mu}$, 
the poles of the integrand of (\ref{eqn: c17}) 
are those of $F_{\mu}$, and 
located at $z_j = x^{-m+j+1}z$. 
For $\mu \geqq 2$, changing the 
contour for $ z_1 $ into 
$ \{ |z_1| = x^{-m+2}(|z |+(m-1) \varepsilon) \} + 
\{ |z_j-x^{-m+j-1}z |=\varepsilon \} - 
\{ |z_j-x^{-m+j+3}z |=\varepsilon \} $,
the integrals are taken on a contour common to 
all $ \mu $. Then the RHS of (\ref{eqn: c17}) 
reduces to its residue at $ z_1 = x^{-m+2}z $:
\begin{eqnarray*}
&& \sum_{\mu=2}^{m}(-1)^{m-1}
\oint_{C_{\mu}'} \prod_{j=2}^{m-1} \frac{dz_j}{2\pi i z_j}
\eta_{m-1}^* (v_m)
\xi^*_{m-1}(v_{m-1})\cdots \xi_1^*(v_1)
\eta_1^* (v_0) \\
&&\times \mbox{Res}_{v_1=v -\frac{m-2}{2}} 
F_{\mu}^*(v_1,\cdots,v_{m-1})
\frac{dz_1}{z_1} \\
&&= B \sum_{\mu=2}^{m}(-1)^{m-2}
\oint_{C_{\mu}'} \prod_{j=2}^{m-1} \frac{dz_j}{2\pi i z_j}
\eta_{m-1}^*(v_m)
\xi^*_{m-1}(v_{m-1})\cdots \xi_1^*(v_1)
\eta_1^*(v_0) \\
&&\times F_{\mu}^{* '}(v_2,\cdots,v_{m-1}) ,
\end{eqnarray*}
where $ v_0 = v - \frac{m-1}{2} , v_1 = v - \frac{m-2}{2} $ and
\[ F_{\mu}^{* '}(v_2,\cdots,v_{m-1}) = \prod_{2\leqq j \leqq m \atop{j \neq \mu}}
\frac{f^*(v_{j-1} - v_j , 1-\pi_{j \mu})}{[\pi_{j \mu}]'} , \]
\[ B = -\mbox{Res}_{v=0}\frac{1}{[v]'}\frac{dz}{z}
= \frac{1}{(x^{2r-2};x^{2r-2})_\infty^3} , \]
and the contour $ C_{\mu}' $ is given by ({\ref{eqn: c18}) 
with $j=2$. Repeating this procedure, we have
\[ 
B^{m-1} 
\eta_{m-1}^*(v_m)
\xi^*_{m-1}(v_{m-1})\cdots \xi_1^*(v_1)
\eta_1^*(v_0) .
\]
\\
Thus Lemma (\ref{eq:m-inv-le}) 
follows from Lemma \ref{lem:prod}. 

For $m=n$, keep (\ref{eq:p-bo=n}) in mind 
and repeat the similar procedure. Then we obtain 
(\ref{eq:m-inv-n}). $\Box$

{\sl Proof of Theorem \ref{thm:inv-II}}. Lemma 
\ref{prop:m-inv} for $m=n$ 
implies (\ref{bp6}). You can also prove 
(\ref{bp7}) in a similar way. $\Box$

{\sl Proof of Theorem \ref{thm:OPE-II}}. 
Set $v_n=v, v_0=v'$ and suppose $\mu >\nu$. 
{}From (\ref{eq:OPE-h*x*}--\ref{eq:h*1l}) 
the product $\bar{\Psi}_{\mu}^{(n-1)}(v)
\Psi_{\nu}^*\left(v'\right)$ is regular 
at $v' = v+\frac{n}{2}$,  
which implies the claim of the theorem 
for $\mu >\nu$. The case $\mu <\nu$ is 
similar. 

Suppose $\mu=\nu$. Then we have 
\begin{equation}
\begin{array}{cl}
&\bar{\Psi}_{\mu}^{(n-1)}(v)
\Psi_{\mu}^*\left(v'\right)
\\
=& (-1)^{n-1} \displaystyle\prod_{j=1}^{n-1}
\oint \frac{dz_j}{2\pi i z_j}
\eta_{n-1}^{*}(v_n)\xi_{n-1}^*(v_{n-1})\cdots
\xi_1^*(v_1)\eta^*_1(v_0)
\prod_{j=1}^{n}f^*(v_{j-1}-v_j,1-\pi_{j \mu}).
\end{array}
\end{equation}
As $v' \to v+\frac{n}{2}$, the countour is pinched.
The limit is calculated by successively taking 
the residues at
$v_{j}=v_{j-1}-\frac{1}{2}$ for $1\leqq j \leqq \mu-1$,
and
$v_{j}=v_{j+1}+\frac{1}{2}$ for $\mu \leqq j \leqq n-1$.
As $z'=z_0 \to x z_{1} \to x^2 z_2 \to \cdots
\to x^n z_n=x^n z$,
the operator part behaves like,
\begin{eqnarray*}
&&\eta_{n-1}^*(v_n)\xi_{n-1}^*(v_{n-1})\cdots
\xi_1^*(v_1)\eta^*_1(v_0)\\
&\to& x^{\frac{1}{2}\frac{r}{r-1}(n-1)}
g_{n-1}^*(x^{-n})
\left(\frac{(x^{2r};x^{2r-2})_\infty}
{(x^{2r-2};x^{2r-2})_\infty}\right)^n
\prod_{j=0}^{n-1}\frac{1}{1-z_{j}/(xz_{j+1})}, 
\end{eqnarray*}
which implies the claim of the theorem. $\Box$

\subsection{Duality relations} 

In this subsection we prove the following 
theorem: 

\begin{thm}{\rm( {\bf Duality}~)} ~~
For $1\leqq \mu \leqq m \leqq n$,
the duality relation is given by
\begin{eqnarray}
\bar{\Psi}_\mu^{(m-1)}(v)=
\prod_{1\leqq j<k\leqq m \atop{j,k \neq \mu}}
{[\pi_{j,k}]'}^{-1}\Psi_\mu^{(m-1)}(v).\label{bp8}
\end{eqnarray}
%where we have set
%\begin{eqnarray}
%c_m=
%(-1)^{\frac{1}{2}(m-1)(m-2)}
%x^{\frac{r}{r-1}\frac{1}{6n}m(m-1)(m-2)}
%\frac{(x^{2r-2};x^{2r-2})_\infty 
%(x^{-2};x^{2r-2})_\infty}
%{g_0^*(x^{-1})g_{m-2}^*(x^{-m+1})}.
%\end{eqnarray}
\label{thm:duality}
\end{thm}

{\sl Proof}. ~~ Define $\tilde{\Psi}_{\mu}^{(m)}(v)$ 
by 
\[
\tilde{\Psi}_{\mu}^{(m)}(v)= 
\prod_{1 \leqq \kappa < \lambda \leqq m+1 \atop{\kappa , 
\lambda \neq \mu}} [\pi_{\kappa , \lambda}]'{}^{-1}\, 
\cdot \Psi_{\mu}^{(m)}(v). 
\]
Let us prove that
\begin{equation}
\tilde{\Psi}_{\mu}^{(m)}(v) = 
\bar{\Psi}_{\mu}^{(m)}(v) 
\label{eqn: c21}
\end{equation}
by induction with respect to $m$. 
The case $m=1$ is trivially true.
First we consider the case $ \mu = m+1 $. 
By performing the cofactor 
expansion for (\ref{df:Psi-m}) we have 
\[
\tilde{\Psi}_{m+1}^{(m)}(v) =
c_m^{-1}
\sum_{\mu=1}^{m} \prod_{1\leqq j \leqq m \atop{j \neq \mu}} 
[\pi_{j \mu}]^{' -1}
\tilde{\Psi}_{\mu}^{(m-1)}(v+\frac{1}{2}) 
\Psi_{\mu}^*(v-\frac{m-1}{2}) .
\]
Because of the assumption of the induction 
and Lemma \ref{prop:m-inv}, we obtain that
$ \tilde{\Psi}_{m+1}^{(m)}(v) = \bar{\Psi}_{m+1}^{(m)}(v) $.
Since $ \tilde{\Psi}_{\mu}^{(m)}(v) $ 
satisfy the same commutation 
relation (\ref{eqn:c11}) as $ \bar{\Psi}_{\mu}^{(m)}(v)$,
taking $ \mu = m+1 $ and calculating
$ (\tilde{\Psi}_{m+1}^{(m)}(v) - 
\bar{\Psi}_{m+1}^{(m)}(v))\Psi_{m+1}^*(v') $,
we have
\begin{equation}
0 =
\sum_{\nu=1}^{m} \Psi_{\nu}^*(v')
(\tilde{\Psi}_{\nu}^{(m)}(v) - 
\bar{\Psi}_{\nu}^{(m)}(v))
\frac{[v-v'-\frac{m+1}{2}-1-
\pi_{\mu \nu}]'}{[v-v'-\frac{m+1}{2}]'}
\prod_{1\leqq \kappa \leqq m1 \atop{\kappa \neq \nu}}
\frac{[1-\pi_{\kappa \mu}]'}{[\pi_{\kappa \nu}]'}
\label{eqn: c22}
\end{equation}
Multiplying $ \bar{\Psi}_{m}^{(n-1)}(v'-\frac{n}{2})$ from 
the left and applying
Theorem \ref{thm:OPE-II}, we get
\[ 0 = \tilde{\Psi}_{m}^{(m)}(v) - \bar{\Psi}_{m}^{(m)}(v) \]
Applying this to (\ref{eqn: c22}) and multiplying 
$\bar{\Psi}_{m-1}^{n-1}(v'-\frac{n}{2})$ from 
the left, we get
\[ 0 = \tilde{\Psi}_{m-1}^{(m)}(v) 
- \bar{\Psi}_{m-1}^{(m)}(v) \]
Repeating this procedure we have (\ref{eqn: c21}).  $\Box$
\\
~\\
{\sl Acknowledgements}~~~
This work was parttly pupported by
the Grant from Ministry of Education, Science, Sports, and
Culture, Japan (11740099).\\

\end{document}